\documentstyle[11pt,twoside,psfig,jltp]{article}
\runninghead{Sergey K. Nemirovskii,  M. Tsubota, and T.
Araki}{Energy Spectrum of the Velocity Field}

\begin{document}

\title{Energy Spectrum of the Random Velocity Field Induced by a Gaussian Vortex
Tangle in HeII}
\author{Sergey K. Nemirovskii,
\address{Institute for Thermophysics, Lavrentyeva,1, 630090,\allowbreak\ Novosibirsk,
Russia} Makoto Tsubota, and Tsunehiko Araki
\address{Department of
Physics, Osaka City University, Osaka, Japan}}

\begin{abstract}
Using the Gaussian model of the vortex tangle (VT) arising in the
turbulent superfluid HeII, we calculate the energy spectrum $E(k)$
of the 3D random velocity field induced by that VT. If the VT is
assumed to be a purely fractal object with Haussdorf dimension
$H_D$, the $E(k)$ is a power-like function $E(k)\propto
k^{-2+H_D}$. A more realistic VT in HeII is a semi-fractal object,
behaving as smooth line for small separations $\Delta \xi \ll R$ (
$\xi $ is the label coordinate, $R$ is mean curvature )and having
a random walk structure for large $\Delta \xi $ with $H_D=2$. For
that case calculations give a spectrum $E(k)$ that is
$k$-independent for $k$ smaller than $1/R$ (but larger than the
inverse size of the system) and that scales as $k^{-1}$ for larger
$k$. The latter reflects the fact that for small scales a vortex
filament behaves as a smooth line. Our results agree with recent
numerical simulations.
\end{abstract}
\maketitle
\section{INTRODUCTION AND SCIENTIFIC BACKGROUND}
The idea that the properties of classical hydrodynamic turbulence
can be described in terms of chaotically moving thin vortex tubes
appeared some time  ago (for a discussion and references see book
by Frisch \cite{Frisch}). Some additional arguments in favor of
that idea would be the recent direct numerical simulations of
flowing liquids with the Reynolds number of order 100. It was
shown that the vorticity field consists of a chaotic set of vortex
tubes resembling the picture appearing in numerical simulations of
the chaotic dynamics of quantized vortex filament in HeII
\cite{Schwarz}$^-$\cite{TAN2000} . Of course the visual similarity
cannot be considered as a confirmation of the equivalence of both
pictures, therefore some quantitative analysis is required to
check whether chaotically moving quantized vortex filaments do
induce a $3D$ velocity field that possesses statistical properties
similar to that of classical turbulence.

In the paper we calculate the spectral density $E(k)$ of the $3D$
random velocity field induced by a chaotic set of quantized vortex
loops. The spectrum $E(k)$ is one of the key characteristics of
turbulent motion. To calculate it we use the Gaussian model of a
VT elaborated by one of authors earlier \cite{N98}. We calculated
spectral density $E(k)$ for the cases of pure fractal lines with
Haussdorf dimension $H_D$ as well as for the more realistic case
of a VT in superfluid turbulent HeII.

\section{GENERAL RELATIONS}

The average kinetic energy of the flow induced by a vortex loop
can be evaluated as follows (see\cite{N98} for details):
\begin{equation}
E=\left\langle \int \frac{{\rho }_s{\bf v}_s^2}2\;d^3{\bf
r}\right\rangle =\int\limits_{{\bf k}}\frac{d^3{\bf k}}{4\pi
k^2}\left\langle 2\pi {\rho }_s{ \kappa
}^2\int\limits_0^L\int\limits_0^L{\bf s}^{\prime }(\xi _1){\bf s}
^{\prime }(\xi _2)\;e^{i{\bf k(s}(\xi _1)-{\bf s}(\xi _2))}d\xi
_1d\xi _2\right\rangle.  \label{E(k)}
\end{equation}
Here ${\bf s}(\xi )$ describes the vortex line position
parameterized by the arclength $ \xi $, running from $0$ to the
length of line $L$, ${\bf s}^{\prime }(\xi )$ denotes the
derivative with respect to arclength along the line (the tangent
vector). Brackets $\left\langle {}\right\rangle $ imply an
averaging over all possible vortex loop configurations. Clearly,
the integrand within the brackets  $ \left\langle {}\right\rangle
$ in (\ref{E(k)}) is just the distribution of energy $dE/dk$
$=4\pi k^2dE/d^3{\bf k}$ in a spherically normalized ${\bf k}$
-space.

Of course relation (\ref{E(k)}) is just a mathematical identity -
the real physical meaning is hidden behind the averaging
procedure. To fulfil an averaging procedure one needs knowledge of
the statistics of a vortex loop. The latter should be taken from
the appropriate stochastic theory of chaotic vortex filaments,
which is unknown at present. Another way is to use numerical
or/and experimental data. The latter gave rise to the Gaussian
model of VT\cite{N98}. In that theory the statistics of a vortex
filament is supposed to be a Gaussian. That supposition is quite
natural if one realizes that a real vortex loop is formed as a
result of numerous reconnections. Therefore the remote vortex line
elements are weakly coupled. The only correlation between them
appears due to propagating Kelvin waves. Then a step-by-step
traversal along a line is a nearly Markovian process.
Correspondingly, statistics of such filaments should be Gaussian
with a sharply decreasing correlation function $N(\xi _1-\xi
_2)=\left\langle {\bf s }^{\prime }(\xi _1){\bf s}^{\prime }(\xi
_2)\right\rangle $ between two remote tangent vectors.

For Gaussian statistics the probability for the line to have some
particular configuration $\{{\bf s}(\xi )\}$ is expressed by the
path integral containing the exponent of the bilinear combination
of functions ${\bf s}(\xi ).$ In practice, to calculate various
averages it is convenient to work with the characteristic
(generating) functional (CF), which is defined as a following
average:

\begin{equation}
W(\{{\bf P}(\xi )\})\;\stackrel{def}{=}\left\langle \exp \left(
i\int\limits_0^L{\bf P}(\xi ){\bf s}^{\prime }(\xi )\ d{\ \xi }\right)
\right\rangle \;.  \label{CFdefinition}
\end{equation}
Since the probability functional is a Gaussian one, calculation of
the CF can be readily made by the full square procedure to give
the result (here we consider only the isotropic case)
\begin{equation}
W(\{{\bf P}(\xi )\})=\exp \left( -\int\limits_0^L\int\limits_0^L\;d\xi
^{\prime }d\xi ^{\prime \prime }{\bf P}{(\xi }^{\prime }{)}N(\xi ^{\prime
}-\xi ^{\prime \prime }){\bf P}{(}\xi ^{\prime \prime }{)}\right) .
\label{W(P'G)}
\end{equation}
From the definition (\ref{CFdefinition}) of the CF it follows that
the function $N(\xi ^{\prime }-\xi ^{\prime \prime })$ coincides
with the correlation function $\left\langle {\bf s}_\alpha
^{\prime }(\xi _1){\bf s} _\alpha ^{\prime }(\xi _j)\right\rangle
$ between tangent vectors. Combining (\ref{E(k)}) and
(\ref{CFdefinition}) one concludes that the spectral density
$E(k)$ is expressed via CF as follows (see\cite{N98} for details):

\begin{equation}
E(k)=\left. 2\pi {\rho }_s{\kappa
}^2\int\limits_0^L\int\limits_0^Ld{\xi } _1d\xi
_2\;\;\frac{{\delta }^2W}{i\delta {\bf P}^\alpha (\xi _1)\;i\delta
{\bf P}^\alpha (\xi _2)}\right| _{{\bf P}_i(\xi _i)\;=\;{\bf
k}\theta (\xi -\xi _1)\theta (\xi _2-\xi )}.  \label{E(W)}
\end{equation}
Here $\theta (\xi )$ is a unit-step function. We will use relation
(\ref {E(W)}) to calculate the spectral density of energy for
various $N(\xi ^{\prime }-\xi ^{\prime \prime }).$

Before doing this let us discuss some fractal properties of vortex
loops. Let us calculate the quantity $D^2=\langle ({\bf s}^\alpha
(\xi _j)-{\bf s}^\alpha (0))^2\rangle $, which is an averaged
squared distance between initial (arbitrary) point of the curve
${\bf s}(0)$ and the points ${\bf s}(\xi _j)$. Note that we deal
with the real distance in three dimensional space (not along the
curve!), therefore this consideration concerns the real size of
the vortex loop embedded in $3D$ space. Using (\ref{W(P'G)}), the
quantity $\langle ( {\bf s}^\alpha (\xi _j)-{\bf s}^\alpha
(0))^2\rangle $ can be evaluated from relations:
\begin{equation}
D^2=\;\int_0^{\xi _J}\;\int_0^{\xi _j}\;d\xi _1\;d\xi _2\langle
{\bf s} _\alpha ^{\prime }(\xi _1){\bf s}_\alpha ^{\prime }(\xi
_2)\rangle \;=\;\int_0^{\xi _j}\;\int_0^{\xi _j}\;d\xi _1\;d\xi
_2N(\xi _1-\xi _2). \label{size1}
\end{equation}

Let us assume that $N(\xi _1-\xi _2)$ is a power-law function
$N(\xi _1-\xi _2)=N*(\xi _1-\xi _2)^\lambda $. Here $N$ is some
constant with the proper dimension. Then the integral in the rhs
of (\ref{size1}) is evaluated as $ \propto \xi _j^{\lambda +2}.$
That implies that the length $L$ of the curve increases with its
$3D$ size $D$ as $L\propto D^{2/(\lambda +2)}$. The latter implies
that the average loop is a fractal object having the Hausdorf
dimension (HD) $H_D=2/(\lambda +2)$. For the example of a pure
random walk chain with the ''effective step'' $l_0$, the function
$N(\xi _1-\xi _2)=l_0\delta (\xi _1-\xi _2)$, and the quantity
$\lambda $ are equal (formally) to $-1$, $H_D=2$, and the length
grows as $L\propto D^2$. For a set of smooth lines $\lambda =0$,
and $H_D=1$. Vortex loops in the superfluid turbulent HeII are
also described by Gaussian statistics. However the function $N(\xi
^{\prime }-\xi ^{\prime \prime })$ is not a power-law function
(see \cite{N98}). The small parts of the filament behave as a
smooth line whose ($3D$) lengths are exactly equal to distance $
\xi ^{\prime }\;-\;\xi ^{\prime \prime }$ along the curve. At the
larger distance the filament has a random walk structure with an
''effective step'' equal to the mean radius of curvature $R$. A
good approximation for the quantity $ N(\xi ^{\prime }-\xi
^{\prime \prime })$ could be a function of type $ 1/(1+(\Delta \xi
/R)^2)$.

\section{ENERGY SPECTRUM OF THE VELOCITY FIELD}

Let us calculate first the energy spectrum for the vortex
filaments, whose statistics are  Gaussian,  with the CF
(\ref{W(P'G)}) and the function $ N(\xi ^{\prime }-\xi ^{\prime
\prime })$, which is a pure power-law function $ N(\xi ^{\prime
}-\xi ^{\prime \prime })=N*\Delta \xi ^\lambda $. Substituting
this into (\ref{E(W)}) one obtains an extremely cumbersome
relation which in general should be evaluated numerically.
However, since we intend to establish the connection between the
fractal properties of the vortex filament and the power-law
spectra of 3D velocity field, it is possible to obtain some
analytical results by focusing on counting powers. By use of the
substitution $z^2=k^2L\Delta \xi ^{\lambda +2}$ relation
(\ref{E(W)}) reduces to an expression having the following
structure.

\begin{equation}
E(k)=L\int\limits_0^{kL^{\frac{\lambda +2}2}}\left\{ Ak^2\left(
\frac zk\right) ^{\frac{4+4\lambda }{\lambda +2}}+B\left( \frac
zk\right) ^{\frac{ 2\lambda }{\lambda +2}}\right\} \exp
(-z^2)z^{\frac{-\lambda }{\lambda +2} }k^{\frac{-2}{\lambda +2}}dz
\label{E(k)power}
\end{equation}
Here $A$ and $B$ are some numerical factors, of order of unity,
appearing from calculating of integrals and $E(k)$ is reduced by
factor $1/2\pi { \rho }_s{\kappa }^2$. It is seen that there are
two different regions of the wave vector $k$, namely, large and
small with respect to $1/L^{(\lambda +2)/2}$, which is nothing but
the inverse $3D$ size of the loop. In the region $k\ll
1/L^{(\lambda +2)/2}$, the exponent is approximately unity, the
term with $B$ disappears due to the closeness of the line, and
simple counting of powers shows that $E(k)\propto k^2$. That
distribution of the energy (valid far from the loop) describes an
equipartition law and implies that $dE/d^3{\bf k}$ is constant. In
the opposite case of large $k$ we can take the upper limit of the
integral to be equal to infinity, after that integral is evaluated
as a combination of gamma functions (of constant arguments)
multiplied by $ k^{-(2+2\lambda )/(\lambda +2)}$. That implies
that the spectral density of energy $E(k)\propto k^{-2+H_D}$. That
result was obtained earlier from very qualitative considerations
and is discussed in the book by Frisch\cite{Frisch}.
\begin{figure}[tbp]
\centerline{\psfig{file=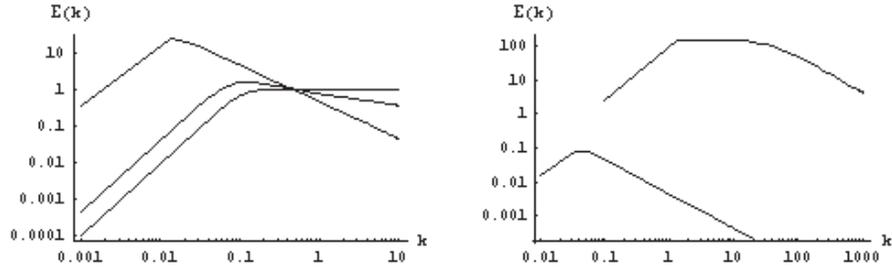,height=1.45in}}
%\framebox[5in]{\rule[1.125in ]{0in}{1.125in}}
%\makebox[5in]{\rule[1.125in]{0in}{1.125in}}
\caption{(a) Energy spectra of  a pure fractal line and (b),
realistic vortex loop in superfluid turbulent HeII (see text). }
\end{figure}

In parallel to a 'counting of powers' we performed a numerical
calculation of the spectra with the help of Mathematica 3.0. In
Fig.1a we depict (in logarithmic scale) three curves corresponding
to the spectral density $E(k)$ for a pure fractal vortex filament
of the length $ L=100$ (all units are arbitrary). The upper curve
(see the region of small $k$) corresponds to the case $\lambda
=0$, i.e. to smooth lines. It can be seen how dependence
$E(k)\propto k^2$ is changed to a  dependence $E(k)\propto k^{-1}$
in the region of $k\sim 100^{-1}$. That is a general result, since
$H_D=1$ and the $3D$ size of the curve is of order of $100$. The
next (middle) curve corresponds to the case $\lambda =-4/5$, i.e.
to the so called self-avoid line. It is seen again that a
$k^2$-dependence is changed to a dependence $E(k)\propto k^{-1/3}$
in the region of $k\sim 100^{-3/5}$. Taking into account that
$H_D=2/(\lambda +2)=5/3$, this behavior agrees with the general
result. The lower curve corresponds to the pure random walk chain
with $N(\xi _1-\xi _2)=l_0\delta (\xi _1-\xi _2).$ Formally
$\lambda =-1$, and $H_D=2.$ Therefore the $k^{2}$-dependance is
changed to $k$-independent spectral density $E(k)\propto k^0$ in
the region of $k\sim 100^{-1/2}$, as it should be.

Let us now consider a VT in HeII. As has been discussed earlier,
the VT in HeII is a semi-fractal object, behaving as a smooth line
for small separation $\Delta \xi \ll R$, and having a random walk
structure for large $ \Delta \xi $ with $H_D=2$ and with the
effective step equal to $R$. Therefore the speculations based on
''counting of powers'' are not applicable directly. However one
can assert that the equipartition distribution of energy $
E(k)\propto k^2$ in the region of $k<<1/\sqrt{LR}$ should take
place for a vortex loop in turbulent HeII. The region of large
wave numbers is divided with some additional value $k\sim 1/R$.
One can suppose that at that point behavior of the spectrum
changes, acquiring the properties of pure fractal with the
corresponding HD in the according  regions. Namely, $E(k)$ should
scale as $k^0$ in region $1/R<<k$, and be approximately
$k$-independent $1/\sqrt{LR} <<k<<1/R$ (provided the latter is
broad enough). Numerical calculations confirm this conjecture. In
Fig. 1b there is depicted (in logarithmic scale) the curves
corresponding to the spectral density $E(k)$ for a vortex loop of
length $ L=100$ with $R=0.01$ and $R=1$. It is clearly seen that
the upper curve ($R=0.01$ ) has three regions with bends at the
points $k_l=1/\sqrt{LR}=1$, and $ k_r=1/R=100.$ In three different
parts we have $E(k)\propto k^2$, $E(k)\propto k^0$ and
$E(k)\propto k^{-1}$, correspondingly. That is in excellent
agreement with qualitative considerations. When the mean radius of
curvature is equal $ R=10,$ the correspondent
$k_l=1/\sqrt{LR}\approx 0.03$, and $k_r=1/R=0.1$. This behavior
agrees with recent numerical simulations\cite{bar01}and our own.

\section{Conclusion}

Using the Gaussian approach for the chaotic vortex filament we
calculate the energy spectrum $E(k)$ of the 3D random velocity
field, induced by that VT. If the VT is assumed to be a purely
fractal object, the $E(k)$ is a power-law function $E(k)\propto
k^{-2+H_D}$. The spectrum of a more realistic VT also agrees with
that formula, if one thinks of VT as a semi-fractal object with a
varying (depending on scale) HD. Let us note that the HD of a
simple noncorrelated vortex loops satisfies the condition
$1<H_D<3$. That implies that Gaussian lines cannot create $3D$
flow, which has the Kolmogorov spectrum $E(k)\propto k^{-5/3}$.
That implies that vortex loops should be strongly correlated
(which, in turn, implies that the local approximation used in
numerical calculations is invalid). Another variant is that the
vortex filaments probably play no special role in turbulent flow.

\section*{ ACKNOWLEDGMENTS}

The work was partly funded by Russian Foundation for Basic Research, Grant N
99-02-16942


\begin{thebibliography}{9}
\bibitem{Frisch}  U. Frisch, {\it Turbulence} (Cambridge University Press,
Cambridge 1996)

\bibitem{Schwarz}  K.W.Schwarz, Phys. Rev. {\bf B38}, 2398 (1988)

\bibitem{NF}  S.K.Nemirovskii and W.Fiszdon, Rev. Mod. Phys., {\bf 67} , 37
(1995)

\bibitem{TAN2000}  Makoto Tsubota, Tsunehiko Araki and Sergey K.
Nemirovskii, Phys Rev. {\bf B62}, 11751 (2000)

\bibitem{N98}  S.K. Nemirovskii, Phys. Rev {\bf B57}, 5792 (1997)

\bibitem{Ed}  M. Doi and S.F. Edwards, {\it The theory of polymer dynamics}
(Clarendon Press, Oxford 1986)

\bibitem{bar01}  D.Kivotides, J.Vassilicos, D.C. Samuels and C.F. Barenghi,
Phys. Rev. Lett., 86, 3080 (2001)
\end{thebibliography}
\end{document}